\documentclass[russian,aps,preprint]{revtex4}

\usepackage[english]{babel}

\usepackage{mathrsfs}
\usepackage{amsmath}
\usepackage{amssymb}
\usepackage{amsfonts}

\begin{document}

\title{Harmonic oscillator coherent states from the orbit theory standpoint}

\author{A. I. Breev }
\email{breev@mail.tsu.ru}
\affiliation{Department of Theoretical Physics, 
             Tomsk State University, 
              Novosobornaya Sq. 1, Tomsk, Russia, 634050}
	
\author{A. V. Shapovalov}
\email{shpv@phys.tsu.ru}
\affiliation{Department of Theoretical Physics, 
             Tomsk State University, 
              Novosobornaya Sq. 1, Tomsk, Russia, 634050}
\affiliation{Laboratory for Theoretical Cosmology, 
             International Centre of Gravity and Cosmos,
             Tomsk State University of Control Systems and Radioelectronics, 
             Lenin ave. 40, Tomsk, Russia, 634050}

\begin{abstract}
	We study the known coherent states of a quantum harmonic oscillator from the 
	standpoint of the original developed noncommutative integration method for 
	linear partial differential equations. The application of the method is based 
	on the symmetry properties of the Schr\"odinger equation and on the orbit 
	geometry of the coadjoint representation of Lie groups. We have shown that 
	analogs of coherent states constructed by the noncommutative integration can be 
	expressed in terms of the solution of a system of differential equations on the 
	Lie group of the oscillatory Lie algebra. The solutions constructed are 
	directly related to irreducible representation of the Lie algebra on the 
	Hilbert space functions on the Lagrangian submanifold to the orbit of the 
	coadjoint representation.
\end{abstract}

\maketitle

\section{Introduction}

The study of exact solutions of the Shrodinger equation for a general
harmonic oscillator has attracted considerable interest in the literature
thanks to the pivotal role of the oscillator in physics. Constructing
exact solutions for a harmonic oscillator based on various ideas and
methods and finding connections between them expands the knowledge
about this fundamental system.

The well-known stationary states of a quantum harmonic oscillator
in the coordinate representation are obtained by separation of variables
in the Schr\"{o}dinger equation with the harmonic oscillator potential. Glauber proposed standart coherent states for a harmonic oscillator
which is the prototype for most of the coherent states \citep{gl1,gl2}.
The coherent states form a very convenient representation for problems
of quantum mechanics. They can be created from the ground state by
displacement operator and can be expanded in terms of the harmonic
oscillator Hamiltonian eigenstates. Coherent states are described
in a wealth of superb reference books and papers, e.g. \citep{Manko1979,perelomov,commb2012}. 

An alternative to the separation of variables method is the noncommutative
integration method (NIM) proposed in \citep{SpSh1995} for linear
partial differential equations and developed in \citep{sh2000,Bar2002,BrSh2016,BrSh2020}.
This method essentially uses the symmetry of a differential equation
and its algebra of symmetry operators and allows one to construct
a basis of solutions that, in general, differs from solutions constructed
by separation of variables and from coherent states. The NIM was effectively
used to construct exact solutions to the Schr\"{o}dinger, Klein-Gordon
\citep{SpSh1995,Bar2002} and Dirac \citep{BrSh2020,BrSh2016,BrSh2014} equations, and also for classification
of external fields in equations with symmetries in the Riemannian
spaces of general relativity in \citep{ob1,ob2,ob3,0b4}. 

This paper describes the development and application of the NIM for
solving the Schr\"{o}dinger equation for a quantum harmonic oscillator,
using symmetry in this problem  We will be looking for  the NIM-solutions
which can be regarded as analogues of coherent states in the sence
of \citep{perelomov,Zh2022} in view of the close relation of NIM
with group symmetry of the quantum harmonic oscillator.

The paper is organized as follows. Section \ref{S2} introduces the
basic notation and a special $\lambda$-representation of Lie algebras
necessary for applying the method of noncommutative integration. Section
\ref{S3} considers the Schr\"{o}dinger equation for a harmonic oscillator
and shows that its symmetry algebra in the class of first-order linear
differential operators forms the oscillatory Lie algebra $\mathfrak{g}_{osc}$.
The next section \ref{S4} considers the $\lambda$-representation
of the oscillatory Lie algebra $\mathfrak{g_{osc}}$ and the generalized
Fourier transform on the Lie group $G_{osc}$ of the Lie algebra $\mathfrak{g_{osc}}$.
Section \ref{S5} shows that the Schr\"{o}dinger equation for an oscillator
is equivalent to some right-invariant system of equations on the group
$G$. Integrating this system by noncommutative integration, we obtain
a basis of solutions and compare it with a system of coherent states.
Section \ref{S6} contains some concluding remarks.

\section{$\lambda$-representation of a Lie groups\label{S2}}

The approach for the noncommutative integration methid is based on
a special representation of the Lie algebra $\mathfrak{g}$ constructed
in terms of the orbit method \citep{kirr1978,kirr2004}. 

First, we recall some necessary definitions from the orbit method
that will be used hereinafter. The degenerate Poisson\textendash Lie
bracket, 
\begin{equation}
\left\{ \phi,\psi\right\} (f)=\langle f,\left[d\phi(f),d\psi(f)\right]\rangle=C_{ab}^{c}f_{c}\frac{\partial\phi(f)}{\partial f_{a}}\frac{\partial\psi(f)}{\partial f_{b}},\quad\phi,\psi\in C^{\infty}(\mathfrak{g}^{*}),\label{pl1-1}
\end{equation}
endows the space $\mathfrak{g}^{*}$ with a Poisson structure~\citep{kirr1978}.
Here, $f_{a}$ are the coordinates of a linear functional $f=f_{a}e^{a}\in\mathfrak{g}^{*}$
relative to the dual basis $\left\{ e^{a}\right\} $, $[\cdot,\cdot]$
is a commutator in the Lie algebra $\mathfrak{g}$, $\langle\cdot,\cdot\rangle$
denotes the natural pairing between the spaces $\mathfrak{g}^{*}$
and $\mathfrak{g}$. The number $\mathrm{ind\mathfrak{g}}$ of functionally
independent Casimir functions $K_{\mu}(f)$ with respect to the bracket
(\ref{pl1-1}) is called the \textit{index of the Lie algebra} $\mathfrak{g}$,
$\mu=1,\dots,\mathrm{ind\mathfrak{g}}.$

A coadjoint representation $\mathrm{Ad}^{*}$: $G\times\mathfrak{g}^{*}\rightarrow\mathfrak{g}^{*}$
splits $\mathfrak{g}^{*}$ into coadjoint orbits (K-orbits). The restriction
of the bracket (\ref{pl1-1}) to an orbit is nondegenerate and coincides
with the Poisson bracket generated by the Kirillov symplectic form
$\omega_{\lambda}$~\citep{kirr1978}. Orbits of maximum dimension
$\mathrm{dim}\mathcal{O}^{(0)}=\mathrm{dim}\mathfrak{g}-\mathrm{ind}\mathfrak{g}$
are called \emph{non-degenerate}~\citep{kirr1978}.

Let $\mathcal{O}_{\lambda}$ be a non-degenerate coajoint orbit passing
through a general covector $\lambda\in\mathfrak{g}^{*}$. Locally,
one can always introduce the Darboux coordinates $(p,q)\in P\times Q$
on the orbit $\mathcal{O}_{\lambda}$ in which the Kirillov form $\omega_{\lambda}$
defining a symplectic structure on the coajoint orbits has the canonical
form $\omega_{\lambda}=dp^{a}\wedge dq_{a}$, $a=1,\dots,\mathrm{dim}\,\mathcal{O}_{\lambda}/2$,
and $(p,q)$ are called the canonical coordinates. We assume that
the transition from the local coordinates $f$ on the orbit $\mathcal{O}_{\lambda}$
to the canonical coordinates $(p,q)$ is given if the set of functions
$f_{X}=f_{X}(p,q,\lambda)$ , $X\in\mathfrak{g}^{*}$ is defined in
such a way that 
\begin{align*}
	& f_{X}(0,0,\lambda)=\langle\lambda,X\rangle,\quad\frac{\partial f_{X}}{\partial p_{\mu}}\frac{\partial f_{Y}}{\partial q^{\mu}}-\frac{\partial f_{X}}{\partial p_{\mu}}\frac{\partial f_{Y}}{\partial q^{\mu}}=f_{[X,Y]},\\
	& \mathrm{rank}\left\Vert \frac{\partial f_{a}}{\partial q^{\mu}},\frac{\partial f_{a}}{\partial p_{\mu}}\right\Vert =\mathrm{dim}\,\mathcal{O}_{\lambda},\quad X,Y\in\mathfrak{g}.
\end{align*}
Consider the functions $f_{X}=f_{X}(p,q,\lambda)$ which are linear
in the variables $p_{a}$: 
\begin{gather}
f_{X}(p,q,\lambda)=\alpha_{X}^{a}(q)p_{a}+\chi_{X}(q,\lambda),\quad q\in Q,\quad p\in P.\label{canf}
\end{gather}
Denote by $\mathfrak{g}_{\mathbb{C}}$ a complex extension of the
Lie algebra $\mathfrak{g}$. It was shown in Ref. \citep{sh2000}
that canonical functions (\ref{canf}) can be constructed if for the
functional $\lambda$ there exists a subalgebra $\mathfrak{h}\subset\mathfrak{g}^{\mathbb{C}}$
in the complex extension $\mathfrak{g}^{\mathbb{C}}$ of the Lie algebra
$\mathfrak{g}$ satisfying the conditions:
\begin{equation}
\langle\lambda,[\mathfrak{h},\mathfrak{h}]\rangle=0,\quad\mathrm{dim}\mathfrak{h}=\mathrm{dim}\mathfrak{g}-\frac{1}{2}\mathrm{dim}\mathcal{O}_{\lambda}.\label{defp}
\end{equation}
The subalgebra $\mathfrak{h}$ is called the \emph{polarization} of
the functional $\lambda$. In this case the vector fields $\alpha_{X}(q)=\alpha_{X}^{a}(q)\partial_{q^{a}}$
are infinitesimal generators of a local transformation group $G_{\mathbb{C}}=\exp(\mathfrak{g}_{\mathbb{C}})$
of a partially holomorphic manifold $Q$. Eq. (\ref{defp}) assumes
that the functionals from $\mathfrak{g}^{*}$ can be prolonged to
$\mathfrak{g}^{\mathbb{C}}$ by linearity. Note that for non-degenerate
coajoint orbits there always exist the canonical functions having\textbf{
}the form\textbf{ }(\ref{canf}). 

Let $L_{2}(Q,d\mu(q))$ be a space of complex functions on the manifold
$Q$ with a measure $d\mu(q)$ and inner product given by
\begin{equation}
(\psi_{1},\psi_{2})=\int_{Q}\overline{\psi_{1}(q)}\psi_{2}(q)d\mu(q),\quad d\mu(q)=\rho(q)dq.\label{sp_Q}
\end{equation}
Here $\overline{\psi_{1}(q)}$ denotes the complex conjugate of $\psi_{1}(q)$.
Functions of the space $L_{2}(Q,d\mu(q))$ are square-integrable on
the manifold $Q$. 

The first-order operators 
\begin{eqnarray}
 &  & \ell_{X}(q,\lambda)=\frac{i}{\hbar}f_{X}(-i\hbar\partial_{q},q,\lambda)=\alpha_{X}^{a}(q)\partial_{q^{a}}+\frac{i}{\hbar}\left[\chi_{X}(q,\lambda)+i\hbar\beta_{X}\right],\quad \partial_{q^a} =\frac{\partial}{\partial q^a}\nonumber\\
 &  &K_{\mu}(-i\hbar\ell(q,\lambda))=K_{\mu}(\lambda),\label{lrpr}\\
 &  & [\ell_{X}(q,\partial_{q},\lambda),\ell_{Y}(q,\partial_{q},\lambda)]=\ell_{[X,Y]}(q,\partial_{q},\lambda),\quad\beta_{\overline{\alpha}}=-\frac{1}{2}\mathrm{Tr}\left(\mathrm{\left.ad_{\overline{\alpha}}\right|}_{\mathfrak{h}}\right),\quad X,Y\in\mathfrak{g,}
\end{eqnarray}
realize, by definition, an irreducible $\lambda$-representation of
a Lie algebra $\mathfrak{g}$ in $L_{2}(Q,d\mu(q))$ and are the result
of $qp$-quantization on the coajoint orbit $\mathcal{O}_{\lambda}$
\citep{SpSh1995,sh2000}.

Without loss of generality, we assume that the operators $-i\hbar\ell_{X}(q,\partial_{q},\lambda)$
are Hermitian with respect to the inner product (\ref{sp_Q}).

We define the generalized functions $\mathscr{D}_{qq'}^{\lambda}(g)$
as solutions of the system 
\begin{eqnarray}
 &  & \left(\eta_{X}(g)+\ell_{X}(q,\partial_{q},\lambda)\right)\mathscr{D}_{qq'}^{\lambda}(g)=0,\label{tD}\\
 &  & \left(\xi_{X}(g)+\overline{\ell_{X}(q',\partial_{q'},\lambda)}\right)\mathscr{D}_{qq'}^{\lambda}(g)=0,\quad\mathscr{D}_{qq'}^{\lambda}(e)=\delta(q,q'),
\end{eqnarray}
where $\xi_{X}(g)=(L_{g})_{*}X$ and $\eta_{X}(g)=-(R_{g})_{*}X$
are left- and right- invariant vector fields on a Lie group $G$,
respectively.

The functions $\mathscr{D}_{qq'}^{\lambda}(g)$ provide the lift of
the $\lambda$-representation of the Lie algebra $\mathfrak{g}$ to
the local unitary representation $T^{\lambda}$ of its Lie group $G$,
\begin{equation}
(T_{g}^{\lambda}\psi)(q)=\int_{Q}\psi(q')\mathscr{D}_{qq'}^{\lambda}(g)d\mu(q),\quad\left.\frac{d}{dt}\left(T_{\exp(tX)}^{\lambda}\varphi\right)\right|_{t=0}(q)=\ell_{X}(q,\partial_{q},\lambda)\varphi(q),\label{prr}
\end{equation}
and satisfy the relations 
\begin{gather*}
\mathscr{D}_{qq'}^{\lambda}(g_{1}g_{2})=\int_{Q}\mathscr{D}_{qq''}^{\lambda}(g_{1})\mathscr{D}_{q''q'}^{\lambda}(g_{2})d\mu(q''),\quad\mathscr{D}_{qq'}^{\lambda}(g)=\overline{\mathscr{D}_{q'q}^{\lambda}(g^{-1})},
\end{gather*}
where $g_{1},g_{2}\in G$. The set of generalized functions $\mathscr{D}_{qq'}^{\lambda}(g)$
satisfying the system of Eqs. (\ref{tD}) has the properties of completeness
and orthogonality for a certain choice of the measure $d\mu(\lambda)$
in the parameter space $J$: 
\begin{align}
 & \int_{G}\overline{\mathscr{D}_{\widetilde{q}\widetilde{q}'}^{\widetilde{\lambda}}(g)}\mathscr{D}_{qq'}^{\lambda}(g)d\mu(g)=\delta(q,\tilde{q})\delta(\tilde{q}',q')\delta(\tilde{\lambda},\lambda),\nonumber \\
 & \int_{Q\times Q\times J}\overline{\mathscr{D}_{qq'}^{\lambda}(\tilde{g})}\mathscr{D}_{qq'}^{\lambda}(g)d\mu(q)d\mu(\lambda)=\delta(\tilde{g}g^{-1}),\label{fD}
\end{align}
where $\delta(g)$ is the generalized Dirac delta function with respect
to the right Haar measure $d\mu(g)$ on the Lie group $G$. 

Note that the functions $\mathscr{D}_{qq'}^{\lambda}(g)$ are defined
globally on the Lie group $G$ iff the Kirillov condition of integerness
of the orbit $\mathcal{O}_{\lambda}$ holds \citep{kirr1978}: 
\begin{equation}
\frac{1}{2\pi}\int_{\gamma\in H_{1}(\mathcal{O}_{\lambda})}\omega_{\lambda}=n_{\gamma}\in\mathbb{Z.}.\label{cel_D}
\end{equation}
Here $H_{1}(\mathcal{O}_{\lambda})$ is a one-dimensional homology
group of the stationarity subgroup $G^{\lambda}=\{g\in G\mid Ad_{g}^{*}\lambda=\lambda\}$. 

Let $L_{2}(G,d\mu(g))$ be the space of functions having the form
\begin{equation}
\psi(g)=\int_{Q}\psi(q,q',\lambda)\mathscr{D}_{qq'}^{\lambda}\left(g^{-1}\right)d\mu(q')d\mu(q)d\mu(\lambda).\label{psiLD}
\end{equation}
Here $\psi(g)$ $\in$$L_{2}(G,d\mu(g))$, and the function $\psi(q,q',\lambda)$
with respect to the variables $q$ and $q'$ belongs to the space
$L_{2}(Q,\mathfrak{h},\lambda)$. We consider equality (\ref{psiLD}) as a generalized\textbf{ }Fourier
transform on the Lie group $G.$ From (\ref{fD}) the inverse transform
can be written as 
\begin{equation}
\psi(q,q',\lambda)=\int_{G}\psi^{\lambda}(g)\overline{\mathscr{D}_{qq'}^{\lambda}\left(g^{-1}\right)}d\mu(g).\label{invF}
\end{equation}
It follows from (\ref{psiLD}) and (\ref{invF}) that the action of
the operators $\xi_{X}(g)$ and $\eta_{X}(g)$ on the function $\psi^{\lambda}(g$)
from $L_{2}(G,\lambda,d\mu(g))$ corresponds to the action of the
operators $\overline{\ell_{X}^{\dagger}(q,\partial_{q},\lambda)}$
and $\ell_{X}(q',\partial_{q'},\lambda)$ on the function $\psi(q,q',\lambda)$:
\begin{align}
 & \xi_{X}(g)\psi^{\lambda}(g)\Longleftrightarrow\overline{\ell_{X}^{\dagger}(q,\partial_{q},\lambda)}\psi(q,q',\lambda)\text{,}\nonumber \\
 & \eta_{X}(g)\psi^{\lambda}(g)\Longleftrightarrow\ell_{X}(q',\partial_{q'},\lambda)\psi(q,q',\lambda).\label{dual}
\end{align}
The functions (\ref{psiLD}) are eigenfunctions for the Casimir operators
$K_{\mu}^{(s)}(i\hbar\xi)=K_{\mu}^{(s)}(-i\hbar\eta)$: 
\begin{align*}
 & K_{\mu}^{(s)}(i\hbar\xi)\psi^{\lambda}(g)\Longleftrightarrow\kappa_{\mu}^{(s)}(\lambda)\psi(q,q',\lambda),\\
 & K_{\mu}^{(s)}(-i\hbar\ell(q',\partial_{q'},\lambda))=\kappa_{\mu}^{(s)}(\lambda),\quad\overline{\kappa_{\mu}^{(s)}(\lambda)}=\kappa_{\mu}^{(s)}(\lambda),\quad\lim_{\hbar\rightarrow0}\kappa_{\mu}^{(s)}(\lambda)=\omega_{\mu}^{(s)}(\lambda).
\end{align*}
As a result of the generalized Fourier transform (\ref{psiLD}), the
left and right fields are converted to $\lambda$-representations,
and the Casimir operators become constants.

\section{Symmetry algebra of a quantum harmonic oscillator\label{S3}}

The states of a one-dimensional quantum harmonic oscillator in the
coordinate representation $\hat{x}=x$, $\quad\hat{p}=-i\hbar\partial_{x}$
are described by the wave function $\psi=\psi(t,x)$ which satisfies
the nonstationary Schr\"{o}dinger equation
\begin{equation}
i\hbar\partial_t=\hat{H}\psi,\quad\hat{H}=\frac{\hat{p}^{2}}{2m}+\frac{m\omega^{2}}{2}\hat{x}^{2},\label{eq:01}
\end{equation}
where $m>0$ is the mass of the quantum particle, $\omega>0$ is the
frequency of the harmonic oscillator,and $\hbar$ is Planck's constant.

The well-known wave functions of the harmonic oscillator in terms
of the Hermite polynomials $H_{n}(z)$ are \citep{perelomov} 
\begin{align}
\psi_{n}(t,x) & =\exp\left(-i\frac{E_{n}}{\hbar}t\right)\psi_{n}(x),\nonumber \\
\psi_{n}(x) & =\langle x\mid n\rangle=\left(\frac{m\omega}{\pi\hbar}\right)^{1/4}\exp\left(-\frac{m\omega x^{2}}{2\hbar}\right)H_{n}\left(\sqrt{\frac{m\omega}{\hbar}}x\right),\nonumber \\
E_{n} & =\hbar\omega\left(n+\frac{1}{2}\right),\quad n=0,1,2,\dots.\label{eq:001}
\end{align}
The eigenstates $\left|n\right\rangle $ for the Hamiltonian $\hat{H}$
are called Fock's or number states, $\hat{H}\left|n\right\rangle =E_{n}\left|n\right\rangle $.
The Fock states are orthonormal and form a complete basis such that
any other state of the harmonic oscillator may be written in terms
of them. 

We can define the annihilation and creation operators by the fotmulas
\[
\hat{a}=\frac{1}{\sqrt{2}}\left(\sqrt{\frac{m\omega}{\hbar}}\hat{x}+\frac{i}{\sqrt{\hbar\omega m}}\hat{p}\right),\quad\hat{a}^{\dagger}=\frac{1}{\sqrt{2}}\left(\sqrt{\frac{m\omega}{\hbar}}\hat{x}-\frac{i}{\sqrt{\hbar\omega m}}\hat{p}\right),\quad[\hat{a},\hat{a}^{\dagger}]=1,
\]
respectively. The time dependent coherent states $\left|z,t\right\rangle $
are eigenstates of the annihilation operator $\hat{a}$, 
\begin{align*}
 & \hat{a}\left|z,t\right\rangle =z(t)\left|z,t\right\rangle ,\quad z(t)=ze^{-i\omega t},
\end{align*}
where the eigenvalue of the operator $\hat{a}$ is a complex number
$z(t)$ which is a function of time $t$. The coherent states may
be written as 
\begin{equation}
\left|z,t\right\rangle =e^{-i\omega t/2}e^{-|z(t)|^{2}/2}\sum_{n=0}^{\infty}\frac{z^{n}(t)}{\sqrt{n!}}\left|n\right\rangle ,\quad\langle z,t\mid z,t\rangle=1.\label{eq:z3}
\end{equation}
In the coordinate representation we have
\begin{equation}
\alpha(t,x;z)=\langle x\mid z,t\rangle=\left(\frac{m\omega}{\pi\hbar}\right)^{1/4}\exp\left[-i\frac{\omega t}{2}-\left(\sqrt{\frac{m\omega}{2\hbar}}x-z(t)\right)^{2}+\frac{z(t)^{2}}{2}-\frac{\left|z\right|^{2}}{2}\right].\label{akk}
\end{equation}
The real and imaginary parts of the quantum number $z$ characterize
the mean values of position and momentum operators:
\[
\langle\hat{x}(t)\rangle=\sqrt{\frac{2\hbar}{m\omega}}\mathrm{Re}\left(z(t)\right),\quad\langle\hat{p}(t)\rangle=\sqrt{2m\hbar\omega}\mathrm{Im}\left(z(t)\right).
\]
Eq. (\ref{eq:01}) admits four integrals of motion in the class of
the first-order linear differential operators:
\begin{align}
iX_{1} & =(\hbar\omega)^{-1}\hat{p}_{0},\quad\hat{p}_{0}=i\hbar\partial_{t},\nonumber \\
-iX_{2} & =-i\sqrt{\frac{m\omega\hbar}{2}}\left(e^{i\omega t}\hat{a}-e^{-i\omega t}\hat{a}^{\dagger}\right)=\cos(\omega t)\hat{p}+m\omega x\sin(\omega t),\nonumber \\
-iX_{3} & =-\sqrt{\frac{m\omega\hbar}{2}}\left(e^{i\omega t}\hat{a}+e^{-i\omega t}\hat{a}^{\dagger}\right)=\sin(\omega t)\hat{p}-m\omega x\cos(\omega t),\nonumber \\
-iX_{4} & =m\omega\hbar.\label{eq:02}
\end{align}
These operators form the Lie algebra $\mathfrak{g}_{osc}$ with non-zero
commutation relations
\begin{equation}
\left[X_{1},X_{2}\right]=-X_{3},\quad\left[X_{1},X_{3}\right]=X_{2},\quad\left[X_{2},X_{3}\right]=-X_{4}.\label{eq:03}
\end{equation}
The algebra $\mathfrak{g}_{osc}$ with the commutation relations (\ref{eq:03})
is called the oscillatory Lie algebra. In the next section, we will
construct a special irreducible $\lambda$-representation of this
Lie algebra, which is necessary for solving the equation (\ref{eq:01})
in terms of the noncommutative integration method.

\section{$\lambda$-representation of the oscillatory Lie algebra \label{S4}}

Let $\{e_{a}\}$ be some fixed basis of the Lie algebra $\mathfrak{g}_{osc}$,
$a=1,\dots,4,$ and $[\cdot,\cdot]$ be the commutator in $\mathfrak{g}_{osc}$,
\[
\left[e_{1},e_{2}\right]=-e_{3},\quad\left[e_{1},e_{3}\right]=e_{2},\quad\left[e_{2},e_{3}\right]=-e_{4}.
\]
An arbitrary element $X\in\mathfrak{g}_{osc}$ is determined by its
components $X^{a}$ with respect to the chosen basis, $X=X^{a}e_{a}$.
In turn, an arbitrary element $f\in\mathfrak{g}_{osc}^{*}$ of the
dual space $\mathfrak{g}^{*}$ is determined by the components of
$f_{a}$ with respect to the basis $\{e^{b}\}$ dual to the basis
$\{e_{a}\},f=f_{a}e^{a}$, $\langle e^{b},e_{a}\rangle=\delta_{a}^{b}$.

The Lie algebra $\mathfrak{g}_{osc}$ admits two Casimir functions
\[
	K_{1}(f)=2f_{1}f_{4}+f_{2}^{2}+f_{3}^{2},\quad
	K_{2}(f)=f_{4},\quad f\in\mathfrak{g}_{osc}^{*}.
\]
Non-degenerate orbits of the coadjoint representation (K-orbits) pass
through the parametrized covector $\lambda=(j_{1},0,0,j_{2})$,
\[
\mathcal{O}_{\lambda}=\left\{ K_{1}(f)=2j_{1}j_{2},\quad K_{2}(f)=j_{2},\quad\neg\left(f_{2}=f_{3}=f_{4}=0\right)\right\} .
\]

Denote by $G_{osc}=\exp\mathfrak{g}_{osc}$ the local Lie group of
the Lie algebra $\mathfrak{g}_{osc}$. Let us introduce canonical
coordinates of the second kind, $x=(x_{1},x_{2},x_{3},x_{4})$, on
the group $G_{osc}$ as
\begin{equation}
g(x_{1},x_{2},x_{3},x_{4})=e^{x_{4}e_{4}}e^{x_{3}e_{3}}e^{x_{2}e_{2}}e^{x_{1}e_{1}}\in G_{osc}.\label{gCc}
\end{equation}
The group composition law in the coordinates (\ref{gCc}) has the
form
\begin{align*}
 & g=g(x_{1},x_{2},x_{3},x_{4}),\quad\tilde{g}=\tilde{g}(y_{1},y_{2},y_{3},y_{4}),\\
 & g\tilde{g}=(g\tilde{g})\bigg{[}x_{1}+y_{1},x_{2}+y_{2}\cos x_{1}+y_{3}\sin x_{1},x_{3}+y_{3}\cos x_{1}-y_{2}\sin x_{1},\\
 & x_{4}+y_{4}+x_{2}\left(y_{1}\sin x_{1}-y_{3}\cos x_{1}\right)+y_{2}y_{3}\sin^{2}x_{1}+\frac{y_{2}^{2}-y_{3}^{2}}{4}\sin(2x_{1})\bigg{]}.
\end{align*}
The Lie group $G_{osc}$ acts on itself by the left $L_{g}$ and right
$R_{g}$ shifts. The left-invariant vector fields $\xi_{a}(g)=(L_{g})_{*}e_{a}$
on the group $G_{osc}$ in local coordinates (\ref{gCc}) are 
\begin{align}
\xi_{1} & =\partial_{1},\quad\xi_{4}=\partial_{4},\nonumber \\
\xi_{2} & =\cos x_{1}\partial_{2}-\sin x_{1}\partial_{3}+x_{2}\sin x_{1}\partial_{4},\nonumber \\
\xi_{3} & =\sin x_{1}\partial_{2}+\cos x_{1}\partial_{3}-x_{2}\cos x_{1}\partial_{4}.\label{eq:left}
\end{align}
The right-invariant vector fields $\eta_{a}(g)=-(R_{g})_{*}e_{a}$
are in turn defined by the expressions
\begin{align*}
\eta_{3} & =-\partial_{1},\quad\eta_{4}=-\partial_{4},\\
\eta_{1} & =-\partial_{1}-x_{3}\partial_{2}+x_{2}\partial_{3}+\frac{1}{2}\left(x_{3}^{2}-x_{2}^{2}\right)\partial_{4},\\
\eta_{2} & =-\partial_{2}+x_{3}\partial_{4}.
\end{align*}
The Lie group $G_{osc}$ is unimodular and the Haar measure coincides
with the Lebesgue measure $d\mu(g)=dx_{1}dx_{2}dx_{3}dx_{4}$. Suppose
that the coordinates $x_{i}$ take values in $\mathbb{R}^{1}$. 

There
exists a three-dimensional complex subalgebra $\mathfrak{h=\mathrm{span}}\{e_{1},e_{2}+ie_{3},e_{4}\}$
of the complex extension $\mathfrak{g}_{\mathbb{C}}^{osc}$ of the
algebra $\mathfrak{g}_{osc}$ subject to the functional $\lambda(j)$,
so that $\langle\lambda(j),\left[\mathfrak{h},\mathfrak{h}\right]\rangle=0$.
This subalgebra is a complex polarization corresponding to the linear
functional $\lambda(j)$. This polarization corresponds to the canonical
transition
\begin{align*}
 & f_{1}(q,p,\lambda)=ipq+j_{1},\quad f_{2}(q,p,\lambda)=-\frac{i}{2}p+j_{2}q,\\
 & f_{3}(p,q,\lambda)=\frac{1}{2}p-ij_{2}q,\quad f_{4}(p,q,\lambda)=j_{2}.
\end{align*}
The $\lambda$ -representation operators are of the form
\begin{align}
 & \ell_{1}(q,\partial_{q},\lambda)=i\left[q\partial_{q}-\frac{1}{\hbar}\left(\frac{j_{2}}{2}q^{2}-j_{1}\right)\right],\quad\ell_{2}(q,\partial_{q},\lambda)=-i\left(\partial_{q}-\frac{j_{2}}{\hbar}q\right),\nonumber \\
 & \ell_{3}(q,\partial_{q},\lambda)=\partial_{q},\quad\ell_{4}(q,\partial_{q},\lambda)=\frac{i}{\hbar}j_{2},\quad Q\in\mathbb{C},\nonumber \\
 & K_2(-i\hbar\ell)=-(\hbar-2j_{1})j_{2},\quad K_2(-i\hbar\ell)=j_2 .\label{eq:lpr}
\end{align}
The function space
\[
\mathscr{F}^{\lambda}=\mathrm{span}\left\{ \varphi_{n}(q)=q^{n}\exp\left(\frac{j_{2}}{4\hbar}q^{2}\right)\mid n=0,1,2,\dots\right\} 
\]
is invariant under the $\lambda$-representation operators and is
a Hilbert space with respect to the scalar product (\ref{sp_Q}) with
the measure
\[
d\mu_{j_{2}}(q)=\exp\left[-\frac{j_{2}}{4\hbar}\left(q-\overline{q}\right)^{2}\right]=\exp\left[-\frac{j_{2}}{4\hbar}\left(q-\overline{q}\right)^{2}\right].
\]

The functions of the space $\mathscr{F}^{\lambda}$ are entire analytic
functions of the complex variable $q$. The generalized Dirac function
in the space $\mathcal{\mathscr{F}}^{\lambda}$,
\[
\psi(q)=\int_{Q}\psi(q')\delta_{j_{2}}(q,\overline{q'})d\mu_{j_{2}}(q'),\quad\psi\in\mathscr{\mathcal{\mathscr{F}}}^{\lambda}
\]
 is defined by the expression
\[
\delta_{j_{2}}(q,\overline{q'})=-\frac{j_{2}}{2\pi\hbar}\sum_{n=0}^{\infty}\frac{\left(-j_{2}/2\hbar\right)^{n}}{n!}\varphi_{n}(q)\overline{\varphi_{n}(q')}=-\frac{j_{2}}{2\pi\hbar}\exp\left[\frac{j_{2}}{4\hbar}\left(q-\overline{q'}\right)^{2}\right].
\]
By integrating the system of equations (\ref{tD}), we obtain
\begin{align}
\mathscr{D}_{qq'}^{\lambda}\left(g^{-1}\right) & =U^{\lambda}(q,g)\delta_{j_{2}}\left(qg^{-1},\overline{q'}\right),\nonumber \\
U^{\lambda}(q,g) & =\exp\left\{ -i\frac{j_{1}}{\hbar}x_{1}-i\frac{j_{2}}{\hbar}x_{4}+\frac{j_{2}}{4\hbar}\left[\left(1-e^{-2ix_{1}}\right)q^{2}+2\left(x_{2}-2iqe^{-ix_{1}}\right)x_{2}\right]\right\} ,\label{5.6}
\end{align}
where $qg^{-1}=q\exp(-ix_{1})+ix_{2}-x_{3}$ is the action of the
group $G_{osc}$ on the complex manifold $Q$, which is given by the
generators
\[
\alpha_{1}(q)=iq\partial_{q},\quad\alpha_{2}(q)=-i\partial_{q},\quad\alpha_{3}(q)=\partial_{q},
\]
so we have
\[
\xi_{X}^{\mu}(g)\frac{\partial(qg)^{a}}{\partial g^{\mu}}=\alpha_{X}^{a}(qg),\quad qe=q,\quad X\in\mathfrak{g}.
\]

The representation (\ref{prr}) becomes an induced representation
of the Lie group $G_{osc}$ and, according to (\ref{5.6}), has the
form
\begin{align*}
 & (T_{g^{-1}}^{\lambda}\psi)(q)=U^{\lambda}(q,g)\psi(qg^{-1}),\\
 & U^{\lambda}(q,\tilde{g}g)=U^{\lambda}(q,g)U^{\lambda}(qg^{-1},\tilde{g}),\quad U^{\lambda}(q,e)=1.
\end{align*}

It can be shown that any $\lambda$-representation of the Lie algebra
in the class of the first-order linear partial differential operators
leads to the induced representation of the Lie group constructed in
the framework of the Kirillov orbit method (see Refs. \citep{BrSh2020,sh2000,SpSh1995}).
The relations (\ref{fD}) are satisfied with respect to the measure
\[
d\mu(\lambda)=\frac{j_{2}}{(2\pi\hbar)^3}dj_{1}dj_{2},\quad\int_{J}(\cdot)d\mu(\lambda)=-\int_{-\infty}^{\infty}dj_{1}\int_{-\infty}^{\infty}(\cdot)j_{2}dj_{2}.
\]
The direct Fourier transform (\ref{psiLD}) in the space $L_{2}(G,d\mu(g))$
has the form
\begin{equation}
\Psi(g)=\int_{Q}\psi(q';q,\lambda)\mathscr{D}_{qq'}^{\lambda}\left(g^{-1}\right)d\mu_{j_{2}}(q')d\mu_{j_{2}}(q)d\mu(\lambda),\quad\psi\in\mathscr{F}^{\lambda}.\label{eq:f1}
\end{equation}

For an invariant second-order differential equation on the group $G_{osc}$
written as 
\begin{equation}
H(-i\hbar\eta)\Psi(g)=0,H(f)=A^{ab}f_{a}f_{b}+B^{a}f_{a}+C,\label{eq:h0}
\end{equation}
where $A^{ab}$, $B^{a}$, $C$ are constants, the general solution
is sought in the form (\ref{eq:f1}) in the framework of NIM. Then
we have the reduced equation for the function $\psi(q';q,\lambda)$,
\begin{equation}
H(-i\hbar\ell(q',\partial_{q},\lambda))\psi(q';q,\lambda)=0,\label{eq:r}
\end{equation}
 which is an ordinary differential equation with respect to the independent
variable $q'$. Eq. (\ref{eq:r}) will be called the equation (\ref{eq:h0}) in the
$\lambda$-\textit{representation}, and the transition from (\ref{eq:h0})
to (\ref{eq:r}) will be called the \textit{non-commutative reduction} of Eq. (\ref{eq:h0}).

\section{The Schr\"{o}dinger Equation on the Oscillatory Lie Group \label{S5}}

Let us show that the Schr\"{o}dinger equation (\ref{eq:01}) describing
the quantum harmonic oscillator is equivalent to the following system
of equations on the Lie group $G_{osc}:$
\begin{align}
	& K_{1}(-i\hbar\xi)\Psi(g)=0,\label{eq:1b}\\
	& K_{2}(-i\hbar\xi)\Psi(g)=\hbar m\Psi(g),\label{eq:1a}\\
	& \eta_{3}\Psi(g)=0.\label{eq:1c}
\end{align}
Indeed, the general solution of the Eqs. (\ref{eq:1a})--(\ref{eq:1c})
can be written as
\[
\Psi(g)=\psi\left(\frac{x_{1}}{\omega},\sqrt{\frac{\hbar}{\omega}}x_{2}\right)e^{imx_{4}}.
\]
Substituting $\Psi(g)$ into the second equation (\ref{eq:1b}), we
get the Schr\"{o}dinger equation for the function $\psi(t,x)$ in
the form
\begin{equation}
i\hbar\frac{\partial\psi}{\partial t}=\hat{H}\psi,\quad\psi=\psi(t,x),\quad x_{1}=\omega t,\quad x_{2}=\sqrt{\frac{\omega}{\hbar}}x.\label{eq:1d}
\end{equation}

Thus, we have reduced the Schr\"{o}dinger equation to the system of
Eqs. (\ref{eq:1a})--(\ref{eq:1c}) on the Lie group
$G_{osc}$ for which the set of basic left-invariant vector fields
(\ref{eq:left}) is a set of non-commuting integrals of motion forming
the Lie algebra $\mathfrak{g}_{osc}$.

Let us integrate the system (\ref{eq:1a})\textendash (\ref{eq:1c})
using the NIM. We are looking for a
solution to this system in the form (\ref{eq:f1}). Then we obtain
the non-commutative reduced system of Eqs. for the function $\psi(q';q,\lambda)$
as
\begin{equation}
\left(j_{1}-\frac{\hbar}{2}\right)\psi(q';q,\lambda)=0,\quad\left(j_{2}+\hbar m\right)\psi(q';q,\lambda)=0,\quad\partial_{q'}\psi(q';q,\lambda)=0.\label{7.1}
\end{equation}

The system (\ref{7.1}) says that the quantum harmonic oscillator
corresponds to the orbit of the coadjoint representation $\mathcal{O}_{\sigma}$
of the group $G_{osc}$, which passes through the parameterized covector
$\sigma=(1/2,0,0,-m)$ , and the function $\psi(q';q,\lambda)$ describing
the quantum harmonic oscillator in terms of the $\lambda$-representation
does not depend on variable $q'$. From (\ref{7.1}) we have
\begin{equation}
\psi(q';q,\lambda)=\psi(q)\delta\left(j_{1}-\frac{\hbar}{2}\right)\delta\left(j_{2}+\hbar m\right).\label{7.2}
\end{equation}
Substituting (\ref{7.2}) into (\ref{eq:f1}) yields the general solution
\begin{align}
 & \Psi(g)=\left.\int_{Q}\psi(q)\mathscr{D}_{qq'}^{\lambda}\left(g^{-1}\right)d\mu_{\lambda}(q)d\mu_{\lambda}(q')\right|_{j_{1}=\hbar/2,\,j_{2}=-\hbar m}\nonumber \\
 & =\int_{Q}\psi(q)\mathscr{D}_{q}\left(x_{1},x_{2},x_{4};\frac{\hbar}{2},-\hbar m\right)d\mu_{-m\hbar}(q),\nonumber \\
 & \mathscr{D}_{q}(x_{1},x_{2},x_{4};j_{1};j_{2})=\int_{Q}\mathscr{D}_{qq'}^{\lambda}\left(g^{-1}\right)d\mu_{j_{2}}(q)=U^{\lambda}(q,g).\label{7.3}
\end{align}

Eq. (\ref{eq:f1}) gives the general solution to the system of equations
(\ref{eq:1a})\textendash (\ref{eq:1c}). According to Eq. (\ref{eq:1d})
the general solution of the Schr\"{o}dinger equation is obtained from
(\ref{7.3}) by setting $x_{1}=\omega t$, $x_{2}=x\sqrt{\omega/\hbar}$, $x_{3}=x_{4}=0$.
It is convenient to represent the general solution of the Schr\"{o}dinger
equation as follows. Let us introduce a set of functions
\begin{equation}
\mathscr{D}\left(t,x\mid u;\mu\right)=\omega\frac{\sqrt{\hbar m}}{2\pi}\left(\frac{\omega m}{\pi\hbar}\right)^{1/4}\mathscr{D}_{u\sqrt{\omega\hbar}}\left(\omega t,\sqrt{\frac{\omega}{\hbar}}x,0;\mu\hbar;-m\hbar\right),\label{7.4}
\end{equation}
which satisfies the completeness and orthogonality conditions:
\begin{align*}
 & \int_{\mathbb{R}^{2}}\overline{\mathscr{D}\left(t,x\mid\tilde{u};\tilde{\mu}\right)}\mathscr{D}\left(t,x\mid u;\mu\right)dtdx=\delta(u,\overline{\tilde{u}})\delta(\mu-\tilde{\mu}),\\
 & \int_{-\infty}^{\infty}d\mu\int_{\mathbb{C}^{1}}d\mu(u)\overline{\mathscr{D}\left(\tilde{t},\tilde{x}\mid u;\mu\right)}\mathscr{D}\left(t,x\mid u;\mu\right)=\delta(t-\tilde{t})\delta(x-\tilde{x}),\\
 & \delta(u,\overline{\tilde{u}})=\frac{m\omega\hbar}{2\pi}\exp\left[-\frac{m\omega\hbar}{4}\left(u-\overline{\tilde{u}}\right)^{2}\right],\quad d\mu(u)=\exp\left[\frac{m\omega\hbar}{4}\left(u-\overline{u}\right)^{2}\right].
\end{align*}

Then the general solution of the Schr\"{o}dinger equation, according
to (\ref{7.3}), is written as
\begin{align}
\psi(t,x) & =\int_{\mathbb{C}^{1}}\overline{\varphi(u)}\mathscr{D}\left(t,x\mid u;1/2\right)du,\quad\label{3.5}\\
u & =q/\sqrt{\omega\hbar}\in\mathbb{C}^{1},\quad\varphi\in\mathscr{F}^{\lambda_{\omega}},\quad\lambda_{\omega}=\left(\frac{1}{2},0,0,-m\omega\hbar\right).
\end{align}

Moreover, for the solution norm (\ref{3.5}) we have
\[
\left\Vert \psi\right\Vert ^{2}=\int_{-\infty}^{\infty}\left|\psi(t,x)\right|^{2}dx=\frac{\omega}{2\pi}\int_{-\infty}^{\infty}\left|\varphi(u)\right|^{2}d\mu(u)=\frac{\omega}{2\pi}\left\Vert \varphi\right\Vert _{Q}^{2}.
\]

As a result, using the NIM, we have
found a general solution (\ref{3.5}) of the Schr\"{o}dinger equation
for the quantum harmonic oscillator. We say that this solution describes
the $H$-state of the harmonic oscillator. Let us show that for a given
solution (\ref{3.5}), stationary solutions are obtained, which are
determined from the equation
\begin{equation}
\hat{p}_{0}\psi(t,x)=E\psi(t,x).\label{3.6}
\end{equation}
Substituting (\ref{3.5}) into (\ref{3.6}) by the function $\varphi(u)$,
we obtain the equation
\[
-i\omega\hbar\ell_{1}(u,\partial_{u},\lambda_{\omega})\varphi(u)=E\varphi(u).
\]
From here we get
\[
\varphi(u)=u^{\frac{E}{\omega\hbar}-\frac{1}{2}}e^{-\frac{m\omega\hbar}{4}u^{2}}.
\]
The function $\varphi(u)$ belongs to the space $\mathscr{F}^{\lambda_{\omega}}$
iff $E/\omega\hbar-1/2=n$, and $n$ is an integer. This condition
results in the well-known spectrum of the quantum harmonic oscillator:
$E_{n}=\hbar\omega(n+1/2)$. The corresponding wave functions on the
manifold $Q$ coincide with the basis functions $\varphi_{n}(q)$
up to a normalization factor: 
\begin{align}
\varphi_{n}(u) & =C_{n}u^{n}e^{-\frac{m\omega\hbar}{4}u^{2}},\nonumber \\
C_{n} & =(-i)^{n}\left(\frac{\omega}{2\pi}\left\Vert \varphi_{n}\right\Vert _{Q}^{2}\right)^{-1/2}=\sqrt{\frac{\hbar m}{2^{n}n!}}\left(-m\omega\hbar\right)^{n/2}.\label{eq:55}
\end{align}
 Then (\ref{3.5}) provides the well-known expression for the wave
functions of the harmonic oscillator in terms of the Hermite polynomials
(\ref{eq:001}) as
\[
\psi_{n}(t,x)=\int_{\mathbb{C}^{1}}\overline{\varphi_{n}(u)}\mathscr{D}\left(t,x\mid u;1/2\right)d\mu(u).
\]

Thus, Fock's states $\left|n\right\rangle $ of the harmonic oscillator
in the $\lambda$-representation (\ref{eq:55}) generate the space
$\mathscr{F}^{\lambda_{\omega}}$ in which the $\lambda$-representation
of the oscillatory group acts.

Comparing (\ref{akk}) and (\ref{7.4}), we obtain the relationship
between the $H$-states and the harmonic oscillator coherent states in
the form 
\[
\mathscr{D}\left(t,x\mid u;1/2\right)=\frac{\omega}{2\pi}\sqrt{\hbar m}\,\alpha\left(t,x;i\sqrt{\frac{m\omega\hbar}{2}}e^{-i\omega t}u\right)\exp\left[-\frac{m\omega\hbar}{4}\left(u^{2}-\left|u\right|^{2}\right)\right].
\]
From here we can see that the $H$-solution (\ref{3.5}) is related to
coherent states of the harmonic oscillator, but it differs from the
latter by a constant factor. In bra-ket notation, the solution(\ref{3.5})
can be represented as
\begin{align}
\left|\psi(t)\right\rangle  & =\int_{\mathbb{C}^{1}}du\overline{\varphi(u)}\left|u,t\right\rangle ,\nonumber \\
\left|u,t\right\rangle  & =\frac{\omega}{2\pi}\sqrt{\hbar m}\exp\left[-\frac{m\omega\hbar}{4}\left(u^{2}-\left|u\right|^{2}\right)\right]\left|z,t\right\rangle ,\quad\langle u,t\mid u,t\rangle=\frac{\omega}{2\pi}\delta(u,\overline{u}),\nonumber \\
z(t) & =i\sqrt{\frac{m\omega\hbar}{2}}e^{-i\omega t}u.\label{eq:8}
\end{align}

Here $\left|z,t\right\rangle $ is a coherent state with a wave function
(\ref{akk}), and the wave function (\ref{3.5}) corresponds to the
state $\left|u,t\right\rangle $. Accordingly, for mean values one
can obtain
\begin{align*}
\langle\hat{x}(t)\rangle_{Q} & =\langle u,t\mid\hat{x}(t)\mid u,t\rangle=-m\left(\frac{\omega\hbar}{2\pi}\right)^{2}\exp\left[-\frac{m\omega\hbar}{2}\left(u^{2}-\left|u\right|^{2}\right)\right]\mathrm{Im}\left(e^{-i\omega t}u\right),\\
\langle\hat{p}(t)\rangle_{Q} & =\langle u,t\mid\hat{p}(t)\mid u,t\rangle=\hbar m\left(\frac{\omega}{2\pi}\right)^{2}\sqrt{2m\hbar\omega}\exp\left[-\frac{m\omega\hbar}{2}\left(u^{2}-\left|u\right|^{2}\right)\right]\mathrm{Re}\left(e^{-i\omega t}u\right).
\end{align*}

From (\ref{eq:z3}) it is easy to write out the expansion of H-states
$\left|u,t\right\rangle $ in terms of Fock's states:
\[
\left|u,t\right\rangle =\frac{\omega}{2\pi}\sqrt{\hbar m}\exp\left[-\frac{i}{2}\omega t-\frac{m\omega\hbar}{4}u^{2}\right]\sum_{n=0}^{\infty}\left(\frac{m\omega\hbar}{2}\right)^{n}\frac{(-1)^{n/2}u^{n}}{\sqrt{n!}}e^{-in\omega t}\left|n\right\rangle .
\]

Thus, as the result of applying the NIM to the system of Eqs.
(\ref{eq:1a})\textendash (\ref{eq:1c}), we have obtained the $H$-states (\ref{eq:8})
of the harmonic oscillator, which, up to a normalization factor, coincide
with known coherent states $\mid z,t\rangle$.

\section{Conclusion\label{S6}}

In this paper, we have shown that the oscillatory Lie algebra $\mathfrak{g_{osc}}$
naturally arises as the Lie algebra formed by the symmetry operators
(\ref{eq:02}) of the Schr\"{o}dinger equation, (\ref{eq:01}) and
the Schr\"{o}dinger equation itself for the harmonic oscillator is
equivalent to a system of right-invariant equations on the corresponding
Lie group $G_{osc}$. As a result of the noncommutative integration
of this system, a complete set of solutions (\ref{3.5}) ($H$-solutions)
is found. Moreover, the quantum harmonic oscillator corresponds to
the only non-degenerate orbit $\mathcal{O}_{\sigma}$ of the adjoint
representation of the Lie group $G_{osc}$. It is shown that the Fock
states of the harmonic oscillator in the $\lambda$-representation
form a Hilbert space $\mathscr{F}^{\lambda_{\omega}}$ which is invariant
under the operators of the $\lambda$-representation (\ref{eq:lpr})
constructed along the given orbit. It turns out that the $H$-solutions
are eigenvalues for the annihilation operator $\hat{a}$, and therefore
they differ from the known coherent states of the harmonic oscillator
by a factor that does not depend on $t$ and $x$ (see Eqs. (\ref{eq:8})),
but depends on the complex quantum number $u$.

\section*{Acknowledgements}

The work is supported by Russian Science Foundation, grant No. 19-12-00042.

\end{document}